\documentclass{aa}

\usepackage{graphicx}
\usepackage{txfonts}
\begin{document}

   \title{A magnetic accretion disk-outflow model for state transition in X-ray binaries}


   \author{Xinwu Cao
          \inst{1,2,3}\fnmsep\thanks{Email: xwcao@zju.edu.cn},
          Bei You \inst{4,5}
\and Zhen Yan\inst{6}          }
   \institute{$^{1}$ Zhejiang Institute of Modern Physics, Department of Physics, Zhejiang University, 38 Zheda Road, Hangzhou 310027, China\\
              $^{2}$ Shanghai Astronomical Observatory, Chinese Academy of Sciences, 80 Nandan Road, Shanghai, 200030, China\\
              $^{3}$ Key Laboratory of Radio Astronomy, Chinese Academy of Sciences,
	210008 Nanjing, China\\
$^{4}$ School of Physics and Technology, Wuhan University, Wuhan 430072, China\\
$^{5}$ Astronomical Center, Wuhan University, Wuhan 430072, China\\
$^{6}$ Shanghai Astronomical Observatory and Key Laboratory for Research Galaxies and Cosmology, Chinese Academy of Sciences, 80 Nandan Road, Shanghai 200030, China
             }

   \date{Received on June 30, 2021; accepted on August 20, 2021}


  \abstract
   {The hard to soft state transition of the outbursts in X-ray binaries (XRBs) is triggered by the rising of the mass accretion rate due to the disk instability. The hard X-ray transition luminosity is found to be tightly correlated to the soft X-ray peak luminosity in the soft state, the physical origin of which is still a mystery. }
   {In order to explain the observed correlation between the hard X-ray transition luminosity and the soft X-ray peak luminosity in the soft state, we construct a magnetic disk-outflow model for the state transition in XRBs.}
   {We assume that the large-scale magnetic field in the outer thin disk is formed  through inverse cascade of small-scale dynamo generated field, and it is then advected by the inner advection dominated accretion flow (ADAF), which accelerates a fraction of the gas into the outflows. The transition luminosity of an ADAF driven by such magnetic outflows varying with the mass accretion rate of the outer disk is calculated.  }
   {During the outbursts, the heating front moves inwards, and the field strength at the heating front of the outer disk is proportional to the accretion rate of the disk. Much angular angular momentum of the inner ADAF is carried away by the outflows for a stronger magnetic field, which leads to a high radial velocity of the ADAF. This makes the critical mass accretion rate of the ADAF increases with the field strength, and it therefore leads to a correlation between transition luminosity and the peak luminosity in the thermal state. We found that the values of the viscosity parameter $\alpha$ of the neutron star XRBs are systematically higher for those of the black hole (BH) XRBs ($\alpha\sim 0.05-0.15$ for BHs, and $\alpha\sim 0.15-0.4$ for neutron stars). {Our model predicts the transition luminosity may be higher than the peak luminosity provided $\alpha$ is sufficiently high, which is able to explain a substantial fraction of outbursts in BHXRBs not reaching the thermally dominant accretion state.}}
  {}


   \maketitle
%

\section{Introduction}\label{intro}


Variability of X-ray binaries (XRBs) exhibits transitions between a thermal state with a
substantial thermal component in the continuum spectra and a
low/hard state dominated by non-thermal emission. It is widely accepted that the states of XRBs correspond to different accretion modes. Hot accretion flows are present in low/hard state, while the black holes (BHs) or neutron stars (NS) are surrounded by geometrically thin accretion disks in the thermal state of XRBs
\citep*[e.g.,][]{1996A&A...314..813L,1997ApJ...489..865E,2000A&A...354L..67M,2004MNRAS.351..791Z,2008ApJ...682..212W,2010LNP...794.....B,2014ApJ...788...52C,2015MNRAS.453.3447D}.

Outbursts in XRBs are generally believed to be triggered by disk instability, which can explain the basic observational features of the outbursts, such as the observed light curves   \citep*[e.g.,][]{1993adcs.book....6C,1998MNRAS.293L..42K,1999MNRAS.303..139D,2001A&A...373..251D}. In the low/hard state, the mass accretion rate is low, and the inner advection dominated accretion flow (ADAF) connects to an outer thin disk at a certain radius, namely the transition radius \citep*[e.g.,][]{1996A&A...314..813L,1997ApJ...489..865E}. During the outburst, the mass accretion rate rises rapidly, and the ADAF is suppressed to a cold disk when its mass accretion rate surpasses over a critical value \citep*[][]{1994ApJ...428L..13N,1995ApJ...452..710N}. As the inner ADAF shrinks, the disk finally approaches the inner stable circular orbits (ISCOs) of BHs or the surface of NS, which corresponds to the thermal state. After
the accretion rate reaches a peak value, it declines slowly as the mass of the disk is exhausted because the rate of the gas re-fueled from the companion star is substantially lower than the central accretion rate \citep*[e.g.][]{1998MNRAS.293L..42K}. The state transitions have been extensively studied by many observers, which provide rigourous constraints on theoretical models  \citep*[e.g.,][]{1995ApJ...442L..13M,1997ApJ...477L..95Z,2002MNRAS.332..856N,2003MNRAS.338..189M,2004MNRAS.351..791Z,2009ApJ...701.1940Y,2010ApJ...718..620W}.    
The observations of the state transitions give important clues on the accretion mode transitions. It is found that the Eddington scaled luminosity of the transition from the hard state to soft state varies in a rather large range, and furthermore it is closely correlated with the peak Eddington ratio of the thermal state \citep*[][]{2004ApJ...611L.121Y,2007ApJ...663.1309Y,2007ApJ...667.1043Y,2009ApJ...701.1940Y}.  These have challenged the theoretical models, because the accretion mode transition is
solely triggered by dimensionless mass accretion rate as expected by the most of the previous models \citep*[see, e.g.,][for reviews]{1998tbha.conf..148N,2014ARA&A..52..529Y}.

In order to explain the hysteretic state transition observed in XRBs \citep*[e.g.,][]{1995ApJ...442L..13M,1997ApJ...477L..95Z,2002MNRAS.332..856N,2003MNRAS.338..189M,2004MNRAS.351..791Z,2009ApJ...701.1940Y}, \citet{2016ApJ...817...71C} suggested a model of the ADAF with magnetically driven outflows, of which critical mass accretion rate for the ADAF may vary with the large-scale magnetic field strength. In this scenario, magnetically driven outflows may carry away a large amount of angular momentum from the ADAF, and therefore significantly increases the radial velocity of the ADAF, which leads to a high critical mass accretion rate. Thus, the luminosity for the transition from hard to soft state increases with the field strength. We conjecture that, such a large-scale magnetic field may form from the inverse cascade of the small-scale field dynamo generated in the outer disk region \citep*[][]{1996MNRAS.281..219T}, the strength of which increases with the peak mass accretion rate of the disk in the outbursts. We describe the model calculations, and the comparison with the observations in Section \ref{model}. Section \ref{discussion} contains brief discussion of the results.

\vskip 1cm
\section{Model}\label{model}

In the quiescent state of XRBs, the mass accretion rate is very low, and the outer cold disk connects to an inner ADAF at a certain radius \citep*[][]{1997ApJ...489..865E}. If the temperature of the gas approaches to the hydrogen ionization temperature somewhere in the cold disks, the instability is triggered at a certain radius of the disk and it makes the annulus transit to the hot state. Heat diffuses rapidly into the adjacent annuli to stimulate them to make the same transition, which triggers an outburst \citep*[e.g.,][]{2002apa..book.....F}. The mass accretion rate $\dot{M}_{\rm d}$ of the disk region in the hot state is significantly higher than that in quiescent state. When the heating front moves inwards into the ADAF, much gas is fed to the inner ADAF. It is suppressed to a thin disk as its accretion rate being higher than a critical value. The ADAF shrinks with rising accretion rate, and finally the ADAF is completely suppressed to a thin disk. {It is then in the thermal state, and the heating front of the outer disk is still moving inwards. The luminosity of the thermal state increases till the heating front approaches the ISCO, which roughly corresponds to the peak luminosity observed in the XRB after the outburst. In the outburst state the disk is quasi-steady, which is well described by the standard disk model \citep*[][]{1998MNRAS.293L..42K}. {Thus, we can model the outburst with a series of ``steady state'' disk calculations. We assume that the heating front propagates inward at constant $\dot{M}_{\rm d}$.} It means that the peak luminosity of the thermal state can be estimated as $L_{\rm peak}\sim\eta_{\rm rad}\dot{M}_{\rm d}c^2$, which is in fact determined by the properties of the disk in quiescent state.}

It was suggested that the magnetically driven outflows carry away a substantial fraction of the angular momentum of the ADAF, which significantly increases the radial velocity of the ADAF \citep*[see][for the details]{2016ApJ...817...71C}. It leads to a high critical mass accretion rate, below which an ADAF with magnetic outflows can survive. An external field can be substantially enhanced in an ADAF due to its large radial velocity \citep*[][]{2011ApJ...737...94C}. It was therefore suggested that ADAFs with magnetically driven outflows are present in the low/hard states of XRBs \citep*[][]{2016ApJ...817...71C}. The external field dragged by the ADAF is taken as a model ingredient in that work, as its origin is still unclear.

The magnetic field strength at the surface of the companion stars in the low mass X-ray binaries (LMXRBs) is around several Gauss, and the field advection in a thin turbulent disk is rather inefficient \citep*[][]{1994MNRAS.267..235L}. This implies that the field formed through the advection in the outer thin disk of LMXRBs may probably be too weak to be the external field for the inner ADAF \citep*[see the estimate of the field strength required for launching outflows from an ADAF in][or the discussion in Section \ref{mdot_crit_adaf}]{2016ApJ...817...71C}.

Recently, magneto-hydrodynamical (MHD) numerical simulations show that turbulence in a radially extended accretion disk can generate large-scale poloidal magnetic flux in situ \citep*[][]{2020MNRAS.494.3656L}. We conjecture that the large-scale field in the outer thin disks of LMXRBs is generated locally. One of the possible mechanisms is that the large-scale field is produced from the small-scale field created by dynamo processes in the disk through an inverse cascade process \citep*[][]{1992MNRAS.259..604T,1996MNRAS.281..219T}. The large-scale magnetic field generated in this way in the outer disk can be advected inwards by the inner ADAF, which accelerates outflows from the ADAF. The critical mass accretion rate, correspondingly the transition luminosity from hard state to soft state, increases with the strength of the magnetic field threading the ADAF, which is roughly proportional to the external field strength, i.e., the strength of the large-scale field generated in the outer disk. The field strength increases with the peak accretion rate of the disk in the soft state. Therefore, the observed tight correlation between the transition luminosity and the peak luminosity in the soft state of XRBs can be reproduced by this scenario.

\vskip 1cm
\subsection{Critical accretion rate for an ADAF with magnetically driven outflows}\label{mdot_crit_adaf}

The hot ADAF is switched to an optically thick accretion disk when its accretion rate is increased above a critical value $\dot{M}_{\rm crit}$ \citep*[][]{1995ApJ...452..710N,1997ApJ...489..865E}. Letting ion-electron equilibration timescale equal to accretion timescale of an ADAF, the critical mass accretion rate $\dot{M}_{\rm crit}$ is estimated
\citep*[][]{1998tbha.conf..148N}. For ADAFs, the temperatures distributions of ions and electrons are almost independent of the mass accretion rate. The rate of energy transfer rate $q_{\rm ie}=n_{\rm i}n_{\rm e}f(T_{\rm i},T_{\rm e})$, in which the detailed form of $f(T_{\rm i},T_{\rm e})$ is given in many previous works
\citep*[e.g.,][]{1983MNRAS.204.1269S,1995ApJ...452..710N,1998MNRAS.296L..51Z}. Thus, we estimate the ion-electron equilibration timescale as
\begin{equation}
t_{\rm ie}\sim{\frac {u}{q_{\rm ie}}}\propto{\frac 1 \rho}\propto\alpha\dot{M}^{-1},\label{t_ie}
\end{equation}
where $\alpha$ is the viscosity parameter, and the relation $\rho\propto \alpha^{-1}\dot{M}$ for an ADAF is used \citep*[][]{1995ApJ...452..710N}.

The accretion timescale is
\begin{equation}
t_{\rm acc}\sim {\frac {R}{|V_{R}|}}={\frac {R^2}{\alpha c_{\rm s}H}}, \label{t_acc}
\end{equation}
where $H$ is the thickness of the ADAF, the radial velocity of the ADAF,
\begin{equation}
V_R\simeq -{\frac \nu R},\label{v_r}
\end{equation}
and the viscosity $\nu=\alpha c_{\rm s}H$ have been used.

Combing Equations (\ref{t_ie}) and (\ref{t_acc}), the critical mass accretion rate is estimated as
\begin{equation}
\dot{M}_{\rm crit,0}\propto\alpha^2.
\label{mdot_crit}
\end{equation}
Detailed model calculations show
\begin{equation}
\dot{M}_{\rm crit,0}\simeq\xi_{\rm c} \alpha^2\dot{M}_{\rm Edd},\label{mdot_crit2}
\end{equation}
where $\dot{M}_{\rm Edd}=L_{\rm Edd}/0.1c^2$, $\xi_{\rm c}=0.1$ and $1$ for NS and BHs respectively \citep*[][]{1995ApJ...452..710N}.

The structure of the disk will be altered if magnetic outflows carry away a fraction of angular momentum of the gas in the disk
\citep*[e.g.,][]{2006A&A...447..813F,2013ApJ...765..149C,2014ApJ...786....6L,2014ApJ...788...71L}.
In the case of an ADAF with magnetically driven outflows carrying substantial angular momentum away from the accretion flow, its radial velocity will be increased, and therefore the critical mass accretion rate is significantly increased, which has been explored in detail by \citet{2016ApJ...817...71C}. We briefly summarize the calculations below \citep*[see][for the details]{2016ApJ...817...71C}.

The magnetic torque exerted by the outflows on unit area of the disk is
\begin{equation}
T_{\rm m}={\frac {B_zB_{\phi}^{\rm s}}{2\pi}}R,\label{t_m_1}
\end{equation}
where $B_z$ and $B_{\phi}^{\rm s}$ are vertical and azimuthal component of the large scale
magnetic field at the disk surface, respectively.
The radial velocity of an ADAF with magnetic outflows is
\begin{displaymath}
V_R^\prime=V_{R}+V_{R,\rm m}=-\alpha c_{\rm s}{\frac {H}{R}}-{\frac
{2T_{\rm m}}{\Sigma
R\Omega}}=-\alpha c_{\rm s}{\frac {H}{R}}- {\frac {B_zB_{\phi}^{\rm
s}}{\pi\Sigma\Omega}}
\end{displaymath}
\begin{equation}
~~~~~~~~~~~~~~~~~~~~~~~~~~~=V_R\left(1+{\frac {B_zB_{\phi}^{\rm
s}}{\pi\Sigma\Omega}}{\frac {R}{\alpha c_{\rm
s}H}}\right), \label{v_r4}
\end{equation}
which can be re-written as
\begin{equation}
V_R^\prime=V_R\left(1+{\frac {4\xi_\phi}{\tilde{H}\beta\alpha
f_\Omega}}\right)=(1+f_{\rm m})V_R,\label{v_r3}
\end{equation}
where the dimensionless quantities are defined as
\begin{equation}
\xi_\phi=-{\frac {B_{\phi}^{\rm s}}{B_z}}, ~~~\beta={\frac {p_{\rm gas}}{B_z^2/8\pi}},~~~f_{\Omega}={\frac {\Omega}{\Omega_{\rm K}}},~~~ \tilde{H}={\frac H R}, \label{def_quan}
\end{equation}
and
\begin{equation}
f_{\rm m}={\frac {4\xi_\phi}{\tilde{H}\beta\alpha f_\Omega}}.
\label{f_m}
\end{equation}
The ratio $\xi_\phi\la 1$ \citep*[see the detailed discussion
in][]{1999ApJ...512..100L}.

The ion-electron equilibration timescale of an ADAF with magnetic outflows is
\begin{equation}
t_{\rm ie}^\prime\propto\alpha(1+f_{\rm m})\dot{M}^{-1},
\label{t_ie2}
\end{equation}
and its accretion timescale becomes
\begin{equation}
t_{\rm acc}^\prime\sim {\frac {R}{|V_{R}^\prime|}}={\frac
{R^2}{\alpha c_{\rm s} H(1+f_{\rm m})}}, \label{t_acc2}
\end{equation}
Similar to the conventional ADAF case, the
value of $\dot{M}_{\rm crit}$ can be estimated with $t_{\rm
ie}^\prime=t_{\rm acc}^\prime$, which leads to
\begin{equation}
\dot{M}_{\rm crit}=(1+f_{\rm m})^2\dot{M}_{\rm crit,0}=\xi_{\rm c}\alpha^2(1+f_{\rm m})^2\dot{M}_{\rm Edd}, \label{mdot_crit3}
\end{equation}
Here the local structure (ion and electron temperatures) is assumed to be similar to a conventional ADAF without magnetic outflows, which is justified in \citet{2016ApJ...817...71C}.

The gravitational energy dissipation rate from the unit area of the
disk is
\begin{equation}
Q_+={\frac {1}{2}}\nu\Sigma\left(R{\frac {d\Omega}{dR}}\right)^2.
\label{q_plus}
\end{equation}
For a conventional ADAF accreting at the critical rate, its gravitational energy dissipation rate from the unit area of the surface,
\begin{equation}
Q_+={\frac {1}{4\pi}}\left(R{\frac
{d\Omega}{dR}}\right)^2\dot{M}_{\rm crit,0}={\frac
{1}{4\pi}}\left(R{\frac {d\Omega}{dR}}\right)^2\xi_{\rm c}\alpha^2\dot{M}_{\rm Edd}, \label{q_plus2}
\end{equation}
where Equations (\ref{v_r}) and (\ref{mdot_crit2}) are used.

Similarly, the gravitational energy dissipation rate of an ADAF with magnetic outflows is calculated with
\begin{displaymath}
Q_+^\prime={\frac {1}{4\pi}}\left(R{\frac
{d\Omega}{dR}}\right)^2{\frac {\dot{M}_{\rm crit}}{(1+f_{\rm
m})}}
\end{displaymath}
\begin{equation}
={\frac {1}{4\pi}}\left(R{\frac {d\Omega}{dR}}\right)^2(1+f_{\rm m})\dot{M}_{\rm
crit,0}=(1+f_{\rm m})Q_+, \label{q_plus3}
\end{equation}
where Equations (\ref{v_r3}) and (\ref{mdot_crit3}) are substituted into Equation  (\ref{q_plus}).

The radial energy advection is negligible in the ADAF accreting at the critical rate \citep*[e.g.,][]{1998tbha.conf..148N,2014ARA&A..52..529Y}, and therefore its luminosity is
\begin{equation}
L_{\rm ADAF}\sim\int 2\pi RQ_+dR. \label{l_adaf}
\end{equation}
For an ADAF with magnetic outflows, its luminosity can be approximated as
\begin{equation}
L_{\rm ADAF}^{\prime}\sim\int 2\pi RQ_+^\prime dR=(1+f_{\rm m})L_{\rm ADAF}, \label{l_adaf_mag}
\end{equation}
which means the ADAF with magnetic outflows is $\sim f_{\rm m}$ times more
luminous than the conventional ADAF accreting at the critical rate.



\vskip 1cm

\subsection{Large-scale magnetic field generated in the outer disk}\label{b_field}

{Origin of a large-scale magnetic field in a thin accretion disk is still uncertain. One possible way is the advection of an external weak field by the disk, however, it is found that the field advection in a thin viscous disk is quite inefficient due to its small radial velocity [$v_R\propto (H/R)^2$] \citep*[][]{{1989ASSL..156...99V,1994MNRAS.267..235L}}. Alternatively, it is suggested that, if most of the angular momentum of the disk is removed by the
magnetically driven outflows, the radial velocity of the disk is significantly higher than that of a conventional viscous thin disk, and the external weak field can be substantially enhanced in the inner region \citep*[][]{2013ApJ...765..149C,2019ApJ...872..149L}. However, the field strength of the companion stars in LMXRBs is too weak to maintain such a thin disk-outflow system \citep*[see][for the detailed discussion]{2019MNRAS.485.1916C}, while an accretion disk with magnetic outflows may be present in the soft state of Cyg X-3 \citep*[][]{2020MNRAS.492..223C}, which is a high mass X-ray binary (HMXRB) containing a Wolf-Rayet star. The observations show that the field strength of Wolf-Rayet stars can be as high as several hundred Gauss \citep*[][]{2014ApJ...781...73D,2016MNRAS.458.3381H}. For our present work, we mostly focus on LMXRBs, and therefore efficient field advection in the thin disk seems to be quite unlikely.}

Recent MHD numerical simulations show that the large-scale field can be generated through dynamo process, i.e., turbulence in a radially extended accretion disk generates  large-scale poloidal magnetic flux in situ even when starting from a purely toroidal magnetic field \citep*[][]{2020MNRAS.494.3656L}. The formation of large-scale magnetic fields has been studied by numerical simulations \citep*[e.g.,][]{2009ApJ...707..428B,2016MNRAS.460.3488S,2018ApJ...857...34Z,2020MNRAS.494.3656L}, mostly for thick disks or the hot coronal flows above the thin disks. So far as we know, no explicit quantitative relation between the disk properties and the large-scale field strength has been revealed in numerical simulations.

We suggest that dynamo process takes place in the outer thin disk in LMXRBs,
however, the length scale of the field generated in this way is around the disk thickness $H$ \citep*[][]{1992MNRAS.259..604T}. In order to launch outflows from the disk, the poloidal magnetic field threading the disk is required to have length scale of $R$.  \citet{1996MNRAS.281..219T} suggested that the inverse cascade process of the dynamo generated small scale magnetic field may lead to large scale field as $B(\lambda)\propto \lambda^{-1}$, where $B(\lambda)$ is the field strength at length scale of $\lambda$.
The strength of such large-scale field is estimated as
\begin{equation}
B_{\rm pd}\sim {\frac {H_{\rm d}} R}B_{\rm dynamo},\label{b_pd_1}
\end{equation}
where $H_{\rm d}$ is the disk thickness, and $B_{\rm dynamo}$ is the dynamo generated field strength \citep*[][]{1996MNRAS.281..219T,1999ApJ...512..100L,2004MNRAS.348..111K}.
{The dynamo generated scale-scale field strength in a thin accretion disk can be estimated as
\begin{equation}
{\frac {B_{\rm dynamo}^2}{4\pi}}\sim \alpha p,
\label{b_dyn}
\end{equation}
\citep*[see the last one of Eq. 2.11 in][]{1973A&A....24..337S}, which is adopted by \citet{1999ApJ...512..100L} (see Equation 10 in their paper) together with Equation (\ref{b_pd_1}) to estimate the power of jets produced through the BP and BZ mechanisms.
Although this scenario has not been reproduced by MHD simulations so far, it is still a promising mechanism for the large-scale field generation in a thin disk with sufficient solid physical ground. In this work, we employ the same approach as \citet{1999ApJ...512..100L} to estimate the strength of the large-scale magnetic field of the thin disk. }

As suggested by \citet{1998MNRAS.293L..42K}, the hot region of the outer disk in the outburst can be described by quasi-steady disk model. {We note that the outburst being close to the peak can be described quite well by a steady disk model, while the density profiles during the outburst may deviate from the standard disk model \citep*[][]{2001A&A...373..251D}. In this work, we focus on the disk conditions near the transition radius during hard-to-soft state transition when the outer disk extends to a small radius connecting to the ADAF, and it is close to the peak of the outburst. The large-scale field strength is estimated from the local structure of the thin disk at the transition radius, and we believe that the local disk structure can be described by the standard thin disk model fairly well.}

{For a radiation pressure dominant disk, its pressure and disk thickness are given by
\begin{equation}
p\simeq p_{\rm rad}=5.81\times 10^{15}\alpha^{-1}m^{-1}r^{-3/2}~~{\rm g~cm}^{-1}~s^{-2},
\label{p_rad}
\end{equation}
and
\begin{equation}
{\frac {H_{\rm d}}{R}}=29.996 \dot{m}_{\rm d}r^{-1},\label{h_rad}
\end{equation}
where $m=M/M_\odot$, $\dot{m}_{\rm d}=\dot{M}_{\rm d}/\dot{M}_{\rm Edd}$, $\dot{M}_{\rm Edd}=4\pi GM/0.1c\kappa_{\rm T}$, and $r=Rc^2/GM$. Although the radiation pressure dominant disk model can describe the outer disk near the transition radius prior to the hard to soft state transition quite well for most cases, we also consider the gas pressure dominant disk model for those observations with low Eddington scaled peak luminosity.
When the dimensionless mass accretion rate is low, the disk will be gas pressure dominant, and the disk pressure and thickness are
\begin{equation}
p\simeq p_{\rm gas}=6.19\times 10^{18}\alpha^{-9/10}m^{-9/10}\dot{m}_{\rm d}^{4/5}r^{-51/20}~~{\rm g~cm}^{-1}~s^{-2},
\label{p_gas}
\end{equation}
and
\begin{equation}
{\frac {H_{\rm d}}{R}}=2.81\times 10^{-2}\alpha^{-1/10}m^{-1/10}\dot{m}_{\rm d}^{1/5}r^{1/20}.\label{h_gas}
\end{equation}
The threshold of the mass accretion rate between radiation pressure dominant and gas pressure dominant disks is calculated by letting the radiation pressure equal to the gas pressure in the disk, which is
\begin{equation}
\dot{m}_{\rm d,\rm rg}=1.64\times 10^{-4}\alpha^{-1/8}m^{-1/8}r^{21/16},
\label{mdot_pr}
\end{equation}
which depends very weakly on $\alpha$ or $m$. }

Thus, the large-scale field strength of the outer disk is estimated as
\begin{equation}
B_{\rm pd}\sim 8.10 \times 10^{9} m^{-1/2}\dot{m}_{\rm d} r^{-7/4}~{\rm Gauss},\label{b_pd_2}
\end{equation}
{when $\dot{m}_{\rm d}\ge \dot{m}_{\rm d,\rm rg}$, and
\begin{equation}
B_{\rm pd}\sim 2.48\times 10^{8} \alpha^{-1/20}m^{-11/20}\dot{m}_{\rm d}^{3/5} r^{-49/40}~{\rm Gauss},\label{b_pd_gas}
\end{equation}
when $\dot{m}_{\rm d}<\dot{m}_{\rm d,\rm rg}$, where Equations (\ref{b_pd_1}) and (\ref{b_dyn}) are used. }


\subsection{Transition of hard state to soft state}

In the low/hard state of XRBs, it is generally believed that an inner ADAF is surrounded by a thin disk in the outer region \citep*[][]{1996A&A...314..813L,1997ApJ...489..865E}. In the outburst, the heating front passes the transition radius $R_{\rm tr}$ into the ADAF, and the mass accretion rate in the ADAF rises to $\dot{M}_{\rm peak}$ at the order of the thermal timescale \citep*[][]{1998MNRAS.293L..42K}. When the accretion rate becomes higher than the critical value $\dot{M}_{\rm crit}$ given by Equation (\ref{mdot_crit3}), the ADAF is suppressed to a standard disk. The strength of the magnetic field threading the the ADAF at $R_{\rm tr}$ is roughly the same as that of the outer disk accreting at $\dot{M}_{\rm peak}$ close to $R_{\rm tr}$, because even the field threading the ADAF is predominantly maintained by the same currents in the outer disk responsible for the large-scale field threading the disk, which is similar to the BH case considered in \citet{1999ApJ...512..100L}.

\begin{figure}
	\centering
	\includegraphics[width=1.2\columnwidth]{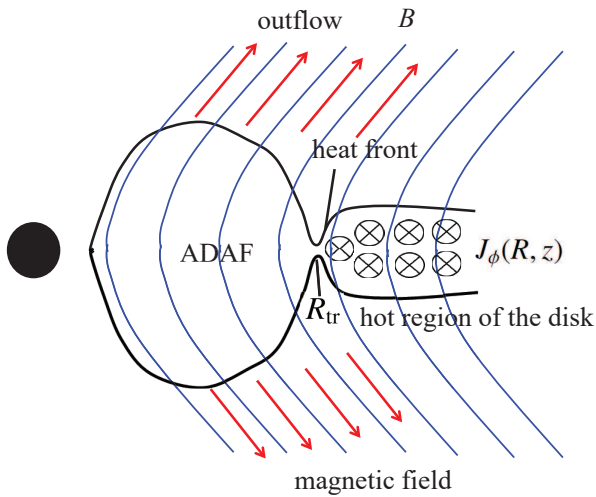}
	\caption{Illustration of the model. The large-scale magnetic field is formed through an inverse cascade process of dynamo generated small-scale field in the outer thin disk. Such a large-scale poloidal field is produced by the azimuthal currents in the disk.  }
	\label{fig_1}
\end{figure}

{Considering the heating front of the outer disk just approaching the ADAF at $R_{\rm tr}$ (see the illustration in Figure \ref{fig_1}), we can estimate the large-scale field strength of the disk very close to transition radius as described in Section \ref{b_field}. Without loss of generality, such a large-scale poloidal magnetic field is maintained by the azimuthal currents of the disk, which means it corresponds to a certain distribution of the currents in the outer disk. The large-scale poloidal field strength of the disk can be calculated with the Biot-Savart law if the current distribution is known \citep*[e.g., see Equation 18 in][]{1994MNRAS.267..235L}. Thus, the strength of the field threading the ADAF at $R_{\rm tr}$ due to the outer disk can be calculated with the same current distribution via the Biot-Savart law. The field strength only depends on the location for a given current distribution, which means that the strength of the field threading the ADAF at $R_{\rm tr}$ is almost same as the large-scale field strength of the outer disk close to the transition radius, because their radial distance is less than the width of heating front during the state transition, which is at the order of the disk thickness \citep*[cf.][]{2002apa..book.....F}. It implies that the ADAF feels the field generated in the outer disk even if the gas of the disk has not been entered the ADAF.
Thus, the strength of the large-scale field threading the ADAF at $r_{\rm tr}$ can be fairly well estimated with Equation (\ref{b_pd_2}) or Equation (\ref{b_pd_gas}).}

For an ADAF with magnetic outflows, its critical mass rate is given by Equation (\ref{mdot_crit3}) (see Section \ref{mdot_crit_adaf}), so the gas pressure of the ADAF accreting at the critical rate is
\begin{equation}
p\simeq 1.74\times 10^{17}\xi_c\alpha(1+f_{\rm m})m^{-1}r_{\rm tr}^{-5/2}\left({\frac H R}\right)^{-1}~~{\rm g~cm}^{-1}~s^{-2}. \label{p_adaf}
\end{equation}
Combining Equations (\ref{def_quan}), (\ref{f_m}), (\ref{b_pd_2}) and (\ref{p_adaf}), we derive
\begin{equation}
f_{\rm m}(1+f_{\rm m})=59.97\xi_{\rm c}^{-1}\alpha^{-2}\dot{m}_{\rm d}^2r_{\rm tr}^{-1},
\label{f_m_2}
\end{equation}
{for a radiation pressure dominant outer disk (i.e., $\dot{m}_{\rm d}\ge \dot{m}_{\rm d,\rm rg}$), or
\begin{equation}
f_{\rm m}(1+f_{\rm m})=5.61\times 10^{-2}\xi_{\rm c}^{-1}\alpha^{-21/10}m^{-1/10}\dot{m}_{\rm d}^{6/5}r_{\rm tr}^{1/20},
\label{f_m_3}
\end{equation}
for a gas pressure dominant disk (i.e., $\dot{m}_{\rm d}<\dot{m}_{\rm d,\rm rg}$). The peak luminosity of the soft state after the outburst is $L_{\rm peak}=\eta_{\rm rad}\dot{M}_{\rm d}c^2$ (see the discussion in the first paragraph of Section \ref{model}). Adopting a conventional value of the radiation efficiency $\eta_{\rm rad}=0.1$, we have $\lambda_{\rm peak}=\dot{m}_{\rm d}$. Solving the above two equations, we derive how the value of $f_{\rm m}$ varies with mass accretion rate $\dot{m}_{\rm d}$ of the outer disk, when the values of the parameters are specified.
}

For the ADAF accreting at the critical rate, the radial energy advection is negligible \citep*[][]{1995ApJ...452..710N}, so the Eddington scaled transition luminosity $\lambda_{\rm tr}\simeq \dot{m}_{\rm crit}=\xi_{\rm c}\alpha^2(1+f_{\rm m})$. {Thus, we obtain the relation of $\lambda_{\rm tr}$ with $\lambda_{\rm peak}$,}
when the values of the parameters $\xi_{\rm c}$, $\alpha$, and $r_{\rm tr}$ are specified. The relations of $\lambda_{\rm peak}$ with $\lambda_{\rm tr}$ are calculated with $\lambda_{\rm tr}=\xi_{\rm c}\alpha^2(1+f_{\rm m})$, which are shown in Figures \ref{fig_2} and \ref{fig_3} for $r_{\rm tr}=20$ and $30$ respectively. In all of the calculations, the approximations $\xi_\phi\sim 1$ and $f_\Omega\sim 1$ are adopted.
{The neutron star mass $m=1.4$ and the black hole mass $m=8$ are adopted in all calculations.}

{When the mass accretion rate in the disk is sufficiently high, a strong magnetic field is generated in the outer radiation pressure dominant disk, which leads to strong outflows.} In this case, $f_{\rm m}\gg 1$, so $1+f_{\rm m}\sim f_{\rm m}$. Substitute $\dot{m}_{\rm d}=\lambda_{\rm peak}$ and $\lambda_{\rm tr}\simeq \xi_{\rm c}\alpha^2(1+f_{\rm m})\sim \xi_{\rm c}\alpha^2f_{\rm m}$ into Equation (\ref{f_m_2}), we obtain
\begin{equation}
\lambda_{\rm peak}\sim 0.13\xi_{\rm c}^{-1/2}\alpha^{-1}r_{\rm tr}^{1/2}\lambda_{\rm tr}. \label{lambda_cor}
\end{equation}
The linear relation (\ref{lambda_cor}) between $\lambda_{\rm tr}$ and  $\lambda_{\rm peak}$ is consistent with the observed correlation derived in \citet{2009ApJ...701.1940Y}.
{We note that the transition luminosity can be higher than the peak luminosity, i.e., $\lambda_{\rm tr}>\lambda_{\rm peak}$, in the case of strong magnetic outflows. For black hole X-ray binaries (BHXRBs), $\xi_{\rm c}=1$, the outbursts would never reach the thermal state, if $\alpha\ga 0.098 r_{\rm tr}^{1/2}$ (e.g., $\alpha\ga 0.44$ for $r_{\rm tr}=20$), while $\alpha\ga 1.4$ is required for $\lambda_{\rm tr}>\lambda_{\rm peak}$ in the case of neutron star X-ray binaries (NSXRBs) ($\xi_{\rm c}=0.1$ and $t_{\rm tr}=20$ are adopted). This can naturally explain the findings that a substantial fraction of outbursts in BHXRBs do not reach the thermally dominant accretion state \citep*[][]{2016ApJS..222...15T}, while no ``hard-only" outburst has been observed in NSXRBs so far as we know. }

\section{Discussion}\label{discussion}

\begin{figure}
	\centering
	\includegraphics[width=0.9\columnwidth]{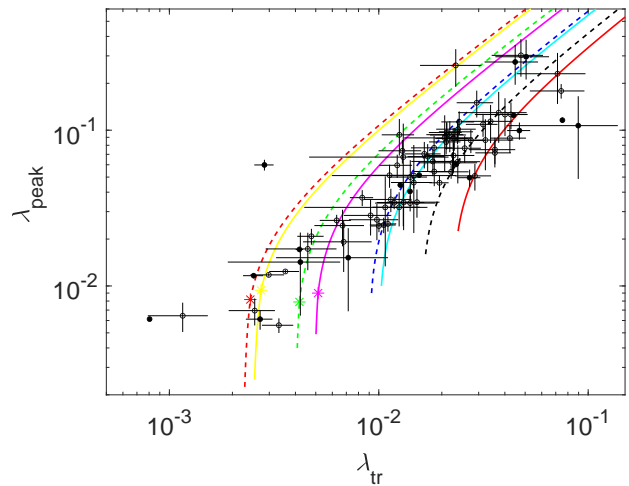}
	\caption{Correlation between $\lambda_{\rm tr}$ and $\lambda_{\rm peak}$. The circles are NSXRBs, while the dots are for the BH counterparts. The data are taken from the literature \citep*[][]{2009ApJ...701.1940Y,2017MNRAS.470.4298Y,2020MNRAS.496.1001C}. The solid lines are the model calculations for BHXRBs (i.e., $\xi_{\rm c}=1$)), while the dashed lines are for NSXRBs (i.e., $\xi_{\rm c}=0.1$). The different colors indicate the results calculated with different values of the viscosity parameter: $\alpha=0.05$ (yellow), 0.07 (magenta), 0.1 (cyan), 0.15 (red), 0.2 (green), 0.3 (blue), and 0.4 (black). The stars indicate the case of the radiation pressure equaling to the gas pressure in the outer disk at $R_{\rm tr}$. The transition radius $r_{\rm tr}=20$ is adopted in all calculations.}
	\label{fig_2}
\end{figure}

\begin{figure}
	\centering
	\includegraphics[width=0.9\columnwidth]{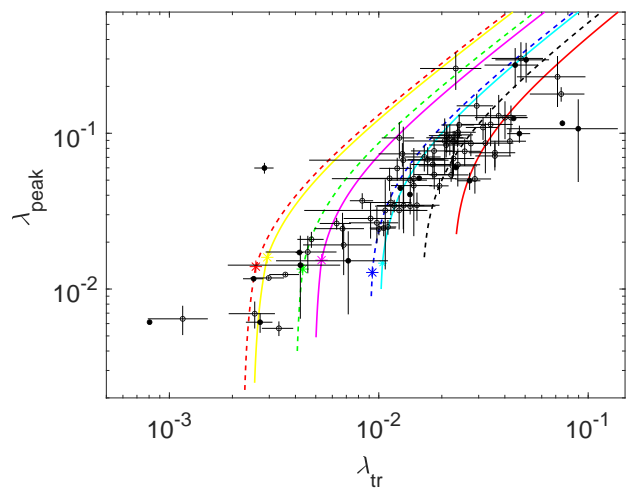}
	\caption{The same as Figure \ref{fig_2}, but $r_{\rm tr}=30$ is adopted.}
	\label{fig_3}
\end{figure}

We estimate of the large-scale magnetic field threading the ADAF at $r_{\rm tr}$ assuming the inverse cascade of dynamo generated field in the outer disk without adding any parameter, which is the same as that employed in \citet{1999ApJ...512..100L}. The ratios $\beta$ of gas to magnetic pressure in the ADAFs are in the range of $\sim 10-10^3$ at the transition radius. With derived field strength, we calculated the angular momentum carried away by the magnetic outflows, and then derived the transition luminosity of the ADAF. We find that the observed correlation between $\lambda_{\rm tr}$ and $\lambda_{\rm peak}$ can be fairly well reproduced by our model calculations without inducing any additional parameter (see Figures \ref{fig_2} and \ref{fig_3}). In our model, strong magnetically driven outflows are present in the outbursts of XRBs, which are indeed consistent with the recent observations \citep*[][]{2018Natur.554...69T}.

{Although the inner truncation radius $r_{\rm tr}$ in the quiescent state can be as large as $\sim 10^4$ \citep*[e.g.,][]{2001A&A...373..251D,2016ApJ...818L...5B},} our model fittings on the observed correlation require the transition radius to be in the range of $\sim 20-30$ (see Figures \ref{fig_2} and \ref{fig_3}), {because all of our calculations are applied for the region at $r_{\rm tr}$ when the state transition takes place. The inner ADAF shrinks with the heating front of the disk moving inwards. As most hard X-ray photons are emitted from the inner region of the ADAF, the transition from hard state to soft state must take place when the transition radius $r_{\rm tr}$ is very close to the ISCO.} Most gravitational energy of the gas in the accretion disk is released in the region with $R\la 2.25 R_{\rm ISCO}$, i.e., $r\la 15$ for a non-rotating BH \citep*[][]{1973A&A....24..337S}. {Thus, it is reasonable to adopt $r_{\rm tr}\sim 20-30$ in the calculations of the state transition.} This is supported by the observational constraints on the transition radius: $r_{\rm tr}\sim 10-10^2$ \citep*[e.g.,][and the references therein]{2015ApJ...813...84G}.

In the outburst, the accretion rate of the ADAF at $r_{\rm tr}$ rises in the thermal timescale $\sim 1/\alpha\Omega_{\rm K}$ when the heating front is approaching the outer radius of the ADAF
\citep*[e.g.,][]{2002apa..book.....F}, {which is roughly at the same order of the accretion timescale of an ADAF without magnetic outflows (see Equation \ref{t_acc}), because the disk thickness of the ADAF $H/R\sim 1$. In the case of the ADAF with strong magnetic outflows, $f_{\rm m}$ is substantially higher than unity, the rising timescale of the accretion rate at $R_{\rm tr}$}is usually longer than the accretion timescale of the ADAF with magnetic outflows (see Equation \ref{t_acc2}). The magnetic coupling between disk and outflows takes place at a much shorter timescale, i.e., the order of the disk dynamical timescale \citep*[][]{2002A&A...385..289C}. These imply that the analyses based on steady model of the ADAF with outflows in this work is a good approximation.

Our model is able to explain the observed correlations naturally both for NSXRBs and BHXRBs, while the values of $\alpha$ are systematically higher for NSXRBs ($\alpha\sim 0.05-0.15$ for BHXRBs, and $\alpha\sim 0.15-0.4$ for NSXRBs). The {tradition ADAF} model shows that the dimensionless critical mass accretion rate for NSXRBs is significantly lower than that of the BH counterparts, because about half of the gravitational energy is released as soft X-ray photons form the NS surface, part of which are injected into the ADAF as soft seed photons inverse Compton scattered by the hot electrons in the ADAF. This increases the radiation efficiency of the ADAF, and therefore reduced the critical mass accretion rate of the ADAF \citep*[see][for the detailed calculations]{1995ApJ...452..710N,1996A&AS..120C.187Y}. However, there is no obvious difference in dimensionless transition luminosity $\lambda_{\rm tr}$ between NSXRBs and BHXRBs (see Figure \ref{fig_2}). Thus, higher values of $\alpha$ are required in our model calculations for NSXRBs than those for BHXRBs. The averages of typical timescales measured with the observations, i.e., rise and decay timescales, and outburst duration, of BHXRBs, are at least two times larger than those of NSXRBs \citep*[][]{2015ApJ...805...87Y}, which is roughly consistent with our results, because the accretion timescale of the disk is proportional to $\alpha^{-1}$, though the physics origin of such difference in $\alpha$ is still unclear.

{In this work, we only calculate the accretion mode transition condition for an ADAF with magnetic outflows at the transition radius. In principle, the critical mass accretion rate for an ADAF with magnetic outflows given by Equation (\ref{mdot_crit3}) is only valid at $R_{\rm tr}$. However, the field advection is very efficient in the ADAF, i.e., the field threading the ADAF at $R_{\rm tr}$ will be dragged inwards, and field strength will be substantially amplified in the central part of the flow \citep*[][]{2011ApJ...737...94C}. It is found that the value of $\beta$ increases  radially in the ADAF \citep*[see Figure 8 in][]{2011ApJ...737...94C}, which implies that the critical mass accretion rate of an ADAF with magnetic outflows will increase with decreasing radius (see Equations \ref{f_m} and \ref{mdot_crit3}). The ADAF is shrinking with the rising of the mass feeding rate of the outer disk. We note that the field advection was calculated only for an ADAF without outflows in \citet{2011ApJ...737...94C}. If the angular momentum carried away by the outflows from the ADAF is properly considered, the field will be more enhanced than an ADAF without outflows, because the radial velocity of the ADAF with outflows is higher than its counterpart without flows. It implies that the whole ADAF can survive if the mass accretion rate is lower than the critical value derived at the transition radius. Our model calculation of the critical mass accretion rate is valid in general for an ADAF with magnetic outflows. The calculation of the global structure of such an ADAF with magnetically driven outflows is desired, and it may help understanding the detailed physics of the hard to soft state transition in XRBs, which is beyond the scope of this work.}



\vskip 1cm

\section*{Acknowledgements}

We are grateful to the referee for his/her insightful comments and suggestions.
We thank Wenfei Yu for very helpful discussion. This work is supported by the NSFC (11773050, 11833007, 12073023, 11903024, U1838103,  U1931203, 11773055, U1838203, and U1938114), the science research grants from the China Manned Space Project with NO. CMS-CSST-2021-A06, the CAS grant QYZDJ-SSWSYS023, and Youth Innovation
Promotion Association of CAS (ids. 2020265).

%


\label{lastpage}
\end{document}